\documentclass[10pt]{article}
\usepackage{psfig,epsf}
\usepackage{a4wide,graphicx}
\usepackage{cite}

\begin{document}

\markboth{J. E. Palomar}{Radiative vector meson decay}

\title{Radiative vector meson decay}

\author{J.~E.~Palomar, L.~Roca, E.~Oset, M.~J.~Vicente~Vacas \\
{\it {\small IFIC, Ed. Institutos de Paterna, Apdo. Oficial
22085, 46071 Valencia, Spain }} \\
{\it {\small joan.palomar@indo.es}}}


\maketitle


\begin{abstract}
We study the radiative $\rho$, $\omega$ and $\phi$
decay into $\pi^0 \pi^0 \gamma$ and  $\pi^0 \eta \gamma $
taking into account mechanisms in which there are two
sequential vector-vector-pseudoscalar  or
axial-vector--vector--pseudoscalar steps followed by the
coupling of a vector meson to the photon, considering the
final state interaction of the two mesons.  Other
mechanisms in which two  kaons are produced through the same
sequential mechanisms or from vector meson decay into two kaons which
 undergo final state interaction leading to the final
pair of pions or $\pi^0 \eta$,
are also considered. 
The results of the parameter free theory, together with the
theoretical uncertainties, are compared with the latest
experimental results at Frascati and Novosibirsk.
\end{abstract}



\section{Introduction}

The radiative decays of vector mesons ($\phi$, $\rho$, $\omega$) into $\pi^0 \pi^0 \gamma$ and 
$\pi^0 \eta \gamma$ have been the subject of intense study
since one can get much information
about the nature of the $\sigma$, $f_0(980)$ and $a_0(980)$  resonances 
from the invariant mass distribution of the two pseudoscalars.  The nature of
these scalar meson 
resonances has generated a large debate, to which new 
light has been brought by the 
claim that  they are dynamically generated from multiple
scattering of pseudoscalars \cite{Oller:1997ti}.

\section{Radiative decays}
\begin{figure}
\centerline{\protect\hbox{
\psfig{file=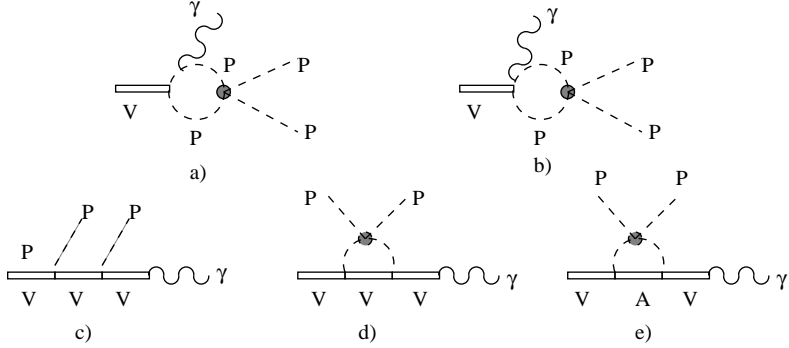,width=0.7\textwidth,silent=}}}
\caption{Diagrams considered. The thick dot in the meson-meson vertices
indicates final state interaction. V denotes a vector meson, A an axial meson
and P a pseudoscalar meson. The
intermediate states in the loops can be $K^+K^-$ or $\pi^+\pi^-$. }
\label{diags}
\end{figure}

In this section we will briefly describe the mechanisms considered to study the decays
and the results obtained. The
 first kind of mechanisms that one can consider are the so called chiral
loops (represented by diagrams a) and b) in figure \ref{diags}), which account for the scalar
 resonances once all the loops have been resummed by using a Bethe-Salpeter equation 
 \cite{Oller:1997ti}. The Lagrangians needed to evaluate these diagrams (appart
 from the ordinary chiral Lagrangians) are the chiral resonance
 Lagrangians:

\begin{center}
\begin{equation}
 {\cal L} = \frac{F_V}{2 \sqrt{2}} < V_{\mu\nu} f^{\mu\nu}_{+} >
+ \frac{iG_V}{\sqrt{2}} < V_{\mu\nu} u^{\mu}u^{\nu} >
\label{resolagr}
\end{equation}
\end{center}

  Another kind of diagrams are the so called vector meson exchange
diagrams (represented by diagrams c) in figure \ref{diags}), considered in some
works  
\cite{Bramon:2001un,Palomar:2002hk} when studying these decays. These diagrams evaluated by means
 of the Lagrangian in eq. \ref{resolagr} and the Lagrangian

\begin{center}
\begin{equation}
{\cal L}_{VVP} = \frac{G}{\sqrt{2}}\epsilon^{\mu \nu \alpha \beta}\langle
\partial_{\mu} V_{\nu} \partial_{\alpha} V_{\beta} P \rangle 
\label{lagr}
\end{equation}
\end{center}

With these two kind of diagrams, and taking into account the $\rho-\omega$
mixing, we find for the $\rho$ and $\omega$
 decays the results shown in table 1.
 
\begin{table}[h]
\caption{Branching ratios due to the different
 contributions to the 
$V \rightarrow P^{0}P^{'0}\gamma$ decays considered.}
{\begin{tabular}{|l|l|l|l|l|}
\hline
BR &  $\rho \rightarrow \pi^{0} \pi^{0}
\gamma$ &  $\rho \rightarrow \pi^{0} \eta \gamma$ & $\omega \rightarrow \pi^{0} \pi^{0} \gamma$ &  $\omega \rightarrow \pi^{0} \eta \gamma$\\
\hline
sequential & $ 1.5\cdot 10^{-5}$ & $6.6\cdot 10^{-10}$ & $4.3\cdot 10^{-5}$ &
$3.0\cdot 10^{-7}$ \\
loops & $1.5\cdot 10^{-5}$ & $5.4 \cdot 10^{-11}$ & $4.3 \cdot 10^{-7}$ &
$2.2\cdot 10^{-9}$ \\
sequential +& & & & \\$\rho$-$\omega$ mixing & not evaluated & $7.5\cdot 10^{-10}$
& $4.8\cdot 10^{-5}$
& $3.4\cdot 10^{-7}$ \\
\hline
Total & $4.2 \cdot 10^{-5}$ & $7.5\cdot 10^{-10}$ & $4.7\cdot 10^{-5}$ &
$3.3\cdot 10^{-7}$ \\ 
\hline
Experiment & $(4.8^{3.4}_{-1.8}\pm 0.2)\times 10^{-5}$  & &
$(7.8\pm 2.7 \pm 2.0) \times 10^{-5}$  & \\
 &   & & $(7.2\pm 2.5) \times 10^{-5}$ & \\
& $(4.1^{+1.0}_{-0.9}\pm 0.3) \times 10^{-5}$  & & $(6.6^{+1.4}_{-0.8}\pm 0.6) \times 10^{-5}$  & \\
 \hline
\end{tabular}}
\end{table}

\noindent As we can see, the chiral loop contribution is only relevant in the
$\rho\rightarrow \pi^{0}\pi^{0}\gamma$ decay.

In the case of the $\phi$ decays\cite{Palomar:2003rb} the sequential vector exchange diagrams
 give a
small contribution since they occur through an OZI-violating $\phi -\omega$
mixing transition. On the other hand, the chiral loops are the dominant
contribution now since the $f_{0}$, $a_{0}$ dominate the amplitudes.

\begin{figure}[tbp]
\centerline{\hbox{\psfig{file=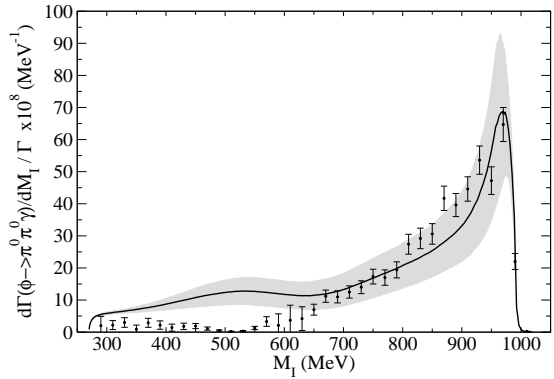,width=8cm,angle=-90}}}
\caption{ Final results for the $\pi^0\pi^0$ invariant mass distribution
for the $\phi\to\pi^0\pi^0\gamma$ decay with the theoretical
error band. Experimental data from Frascati.}
\label{phipi}
\end{figure}

\begin{figure}[tbp]
\centerline{\hbox{\psfig{file=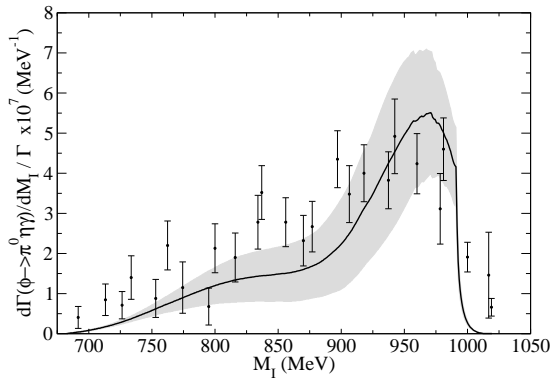,width=8cm,angle=-90}}}
\caption{\rm 
Final results for the $\pi^0\eta$ invariant mass distribution
for the $\phi\to\pi^0\eta\gamma$ decay with the theoretical
error band. }
\label{phieta}
\end{figure}

 The fact
that we have experimental results not only on the integrated branching ratios
but also on the invariant mass distributions makes these decays very appealing.
In fact, when we compare the theoretical calculation provided by the
contributions  of diagrams \ref{diags}a), \ref{diags}b) and \ref{diags}c) with
the experimental results we see that we get a too narrow distribution\footnote{There is also a disagreement in the region around 500 MeV,
 but this is not a
problem since these results have to be reanalized and preliminary results seem
to be in agreement with our predicition.} in the
region of the $f_0$, $a_0$. The chiral loop
contribution is the responsible for the peaks, giving a pole
in the right place but not with a sufficiently large width.
 Therefore, we should take into account more diagrams that have a
sizeable effect but have not been considered in the literature when dealing
with this problem. These are the unitarized sequential exchange with both an
intermediate vector (see
\ref{diags}d)) or axial (see \ref{diags}e)) particle. The evaluation of diagrams \ref{diags}d)
has been done by using the unitarized amplitudes for the pseudoscalar-pseudoscalar interaction
\cite{Oller:1997ti}. As for the diagrams with axial resonances, we have built the lowest order
Lagrangian consistent with chiral symmetry that accounts for these resonances\cite{Roca:2003uk}. This Lagangian
 is given in eq. \ref{eq:LBLA}

\begin{eqnarray}  \label{eq:LBLA}
{\cal L}_{BVP}&=&\tilde{D} <B_{\mu\nu}\{V^{\mu\nu},P\}> \\
\nonumber
{\cal L}_{AVP}&=&i\tilde{F} <A_{\mu\nu}[V^{\mu\nu},P]> 
\end{eqnarray} 

\noindent where $\tilde{D}$ and $\tilde{F}$ are free parameters, $A$ and $B$ are matrices containing the axial fields, and $P$ and $V$ account for the
pseudoscalar and vector mesons, respectively.   To fix the values of the free parameters in
the Lagrangians we have fitted them to ten different decay channels of axials to a vector
and a pseudoscalar. With this we have all the parameters fixed, since the other
parameters of the Lagrangians considered are well known and for the integrals we use the same
cut-off of 1 GeV as the one used in reference \cite{Oller:1997ti} to describe the meson-meson
interaction. The final results for the $\phi\rightarrow\pi^{0}\pi^{0}\gamma$ and 
$\phi\rightarrow\pi^{0}\eta\gamma$ are shown in figures \ref{phipi} and \ref{phieta}.

\section{Conclusions}
As we have seen in the previous section, the chiral loops and the sequential vector meson
exchange mechanisms provide a good description of the $\rho$ and $\omega$ radiative decays.
But in the case of the $\phi$ these two mechanisms are certainly not enough, since the peaks
obtained for the scalar resonances are not wide enough. The novelty of this work is the
parameter-free inclusion of the unitarized sequential vector exchange and the unitarized 
sequential axial vector exchange, which helps to widen the distributions in the resonance
region. The results obtained are in agreement with the experimental data. The apparent 
disagreement in the region of around 500 MeV will be
most likely solved once the experimental data had been reanalized.


\section*{Acknowledgements}

I want to aknowledge prof. Chiang and all the
people from IHEP for their warm hospitality and the wonderful organization. I want also to acknowledge with
sincere gratitude prof. E. Oset who made possible
my participation in the conference.

\end{document}